\documentclass{ragtime} 

\title[Quasinormal ringing of Bardeen spacetime]%
      {Quasinormal ringing of Bardeen spacetime}
\author[D. Ovchinnikov]%    Now let's start the paper title authors:
       {D. Ovchinnikov\at[]{1,a}\\% Termination of authors' block; if
\ins{1}Research Centre for Theoretical Physics and Astrophysics,\splitins[1]
       Institute of Physics, Silesian University in Opava,\splitins[1]                                              
       Bezru\v{c}ovo n\'am.~13, CZ-746\,01 Opava,
       Czech Republic\\% Termination of the first affiliation.
\ins{a}\Email{dmitriy.ovchinnikov@physics.slu.cz}} % This is how to present E-mail.

\coentry{D. Ovchinnikov}%
       {Quasinormal ringing of Bardeen spacetime}

\DeclareMathAlphabet{\pazocal}{OMS}{zplm}{m}{n}

\begin{document}

% Citation of references in abstract should generally be avoided to
% ensure self-consistency of the abstract.  If you do insist on citation(s)
% within the abstract, you should use the \bibentry command, which forces
% the _complete_bibliographic_entry_ to appear in the abstract.
% With the `nonatbib' optional class argument this feature is not available.
\begin{abstract}
We review recent calculations of quasinormal modes and asymptotic tails of the Bardeen spacetime interpreted as a quantum corrected Schwarzschild-like black holes. Massless electromagnetic and Dirac fields and massive scalar field are considered. The first few overtones are much more sensitive to the change of the quantum correction parameter than the fundamental mode, because such correction deforms the black hole geometry near the event horizon. While the asymptotic tails of massless fields are identical to those for the Schwarzschild case, the tails for a massive field differs from the Schwarzschild limit at both intermediate and asymptotic times. 
\end{abstract}

% The key words are to be separated by the N-dash surrounded by spaces,
% the left one non-breakable.  The name concatenation like Kerr--de~Sitter
% should also be typed with N-dash but with no spaces around (compare,
% e.g., Levi-Civita, which is a single person):

% It is good to provide as many as \label's possible, but never start the
% key with a numeral.  This makes problems with pdflatex processing.
\section{Introduction}\label{intro}%%%%%%%%%%%%%%%%%%%%%%%%%%%%%%%%%%%%%%%%

Quasinormal modes of black holes \cite{Kokkotas:1999bd,Nollert:1999ji,Konoplya:2011qq} are a fundamental aspect of black hole physics and gravitational wave astronomy. These modes represent the characteristic oscillations and decay of perturbations around a black hole after an external perturbation, such as a merger or accretion event. These modes are characterized by complex frequencies, which have direct implications for the detection and interpretation of gravitational wave signals from black hole mergers by observatories like LIGO and Virgo \cite{LIGOScientific:2016aoc}. 

Quasinormal modes of the historically first model of the regular black holes given by the Bardeen spacetime, have been extensively studied in a great number of papers (see for instance \cite{Flachi:2012nv, Toshmatov:2015wga,Toshmatov:2019gxg, MahdavianYekta:2019pol,Rincon:2020cos, Lopez:2022uie,Saleh:2018hba} and reference therein).
However, the Bardeen spacetime was considered there mainly as a solution of a specific non-linear electrodynamics \cite{Ayon-Beato:2000mjt} which describes a black hole as a gigantic magnetic monople with zero electric charge \cite{Bronnikov:2000vy}. 

Recently, quasinormal spectrum of the Bardeen spacetime as a quantum corrected neutral black hole metric \cite{Nicolini:2019irw} has been considered in \cite{Konoplya:2023ahd,Bolokhov:2023ruj} with the emphasis to overtones behavior and asymptotic tails. Here we review these results and discuss three interesting phenomena related to the Bardeen black hole spectrum: a) outburst of overtones b) arbitrarily long lived modes of massive fields and c) asymptotic tails.

\section{Bardeen spacetime and the wavelike equations}\label{sec.Bardeen spacetime and the wavelike equations}%%%%%%%%%%%%%%%%%

The spherically symmetirc line element has the form
\begin{equation}
		d s^2 = - f(r) d t^2 + f^{-1}(r) d r^2 + r^2 (d \theta^2 + \sin^2 \theta d \varphi^2),
\end{equation}
where for the Bardeen spacetime, the metric function is
\begin{equation}
f(r) =  1 - \frac{2M r^2}{ ( r^2 + l_{0}^2 ) ^{3/2}}.
\end{equation}
For $l_{0}\neq 0$, the space-time in eq.(5) has horizons  for  $|l_{0}| \leq  4 M / (3 \sqrt{3})$.

The general relativistic equations for the scalar ($\Phi$), electromagnetic ($A_\mu$), and Dirac ($\Upsilon$) fields in a curved spacetime can be written as follows:
%\begin{subequations}\label{coveqs}
%\begin{eqnarray}\label{KGg}
%\frac{1}{\sqrt{-g}}\partial_\mu \left(\sqrt{-g}g^{\mu \nu}\partial_\nu\Phi\right)&=&0,
%\\\label{EmagEq}
%\frac{1}{\sqrt{-g}}\partial_{\mu} \left(F_{\rho\sigma}g^{\rho \nu}g^{\sigma \mu}\sqrt{-g}\right)&=&0\,,
%\\\label{covdirac}
%\gamma^{\alpha} \left( \frac{\partial}{\partial x^{\alpha}} - \Gamma_{\alpha} \right) \Upsilon&=&0.
%\end{eqnarray}
%\end{subequations}
\begin{subequations}\label{coveqs}
\begin{align} \label{KGg}
&\frac{1}{\sqrt{-g}}\partial_\mu \left(\sqrt{-g}g^{\mu \nu}\partial_\nu\Phi\right)=0,
\\\label{EmagEq}
&\frac{1}{\sqrt{-g}}\partial_{\mu} \left(F_{\rho\sigma}g^{\rho \nu}g^{\sigma \mu}\sqrt{-g}\right)=0,
\\\label{covdirac}
&\gamma^{\alpha} \left( \frac{\partial}{\partial x^{\alpha}} - \Gamma_{\alpha} \right) \Upsilon=0.
\end{align}
\end{subequations}
Here $F_{\mu\nu}=\partial_\mu A_\nu-\partial_\nu A_\mu$ is the electromagnetic tensor, $\gamma^{\alpha}$ are gamma matrices and $\Gamma_{\alpha}$ are spin connections in the tetrad formalism.
Using separation of variables, after some algebra the above dynamical equations (\ref{coveqs}) take the wave-like form:
\begin{equation}\label{wave-equation}
\dfrac{d^2 \Psi}{dr_*^2}+(\omega^2-V(r))\Psi=0,
\end{equation}
where the ``tortoise coordinate'' $r_*$ is:
\begin{equation}\label{tortoise}
dr_*\equiv\frac{dr}{f(r)}.
\end{equation}

The effective potentials for the scalar ($s=0$) and electromagnetic ($s=1$) fields can be written in a unified form:
\begin{equation}\label{potentialScalar}
V(r)=f(r) \frac{\ell(\ell+1)}{r^2}+\left(1-s\right)\cdot\frac{f(r)}{r}\frac{d f(r)}{dr},
\end{equation}
where $\ell=s, s+1, s+2, \ldots$ are the multipole numbers.
For the Dirac field ($s=1/2$) the problem is reduced to two iso-spectral effective potentials
\begin{equation}
V_{\pm}(r) = W^2\pm\frac{dW}{dr_*}, \quad W\equiv \left(\ell+\frac{1}{2}\right)\frac{\sqrt{f(r)}}{r}.
\end{equation}
The iso-spectral wave functions can be transformed one into another by the Darboux transformation
\begin{equation}\label{psi}
\Psi_{+}=q \left(W+\dfrac{d}{dr_*}\right) \Psi_{-}, \quad q=const,
\end{equation}
so that it is sufficient to analyze the spectrum of only one of the potentials.

\section{Long lived quasinormal modes and the outburst of overtones}\label{geneqcond}%%%%%%%%%%%%%%%%%%

The boundary conditions for quasinormal modes are purely outgoing wave at infinity and purely incoming wave at the event horizon, so that
\begin{equation}
	\Psi=\left\{\begin{array}{lcr}
	e^{i\omega r_*}, & \textrm{ for } r_* \to +\infty  & \textrm{ (purely outgoing) },\\
	e^{-i\omega r_*}, & \textrm{ for } r_* \to -\infty  & \textrm{ (purely ingoing) },
	\end{array}\right.
	\label{eq:bc}
\end{equation}

In order to find low-lying quasinormal frequencies, the quick and relatively accurate method which was used in \cite{Konoplya:2023ahd,Bolokhov:2023ruj} is the 6th order WKB method \cite{Konoplya:2003ii,Konoplya:2019hlu} with the Pade approximants \cite{Matyjasek:2017psv}. The WKB method was effectively used in a great number of works (see for example \cite{Kodama:2009bf,Onozawa:1995vu,Konoplya:2019hlu} and references therein). In order to find accurate values of overtones with $n>\ell$ the convergent Leaver method was used \cite{Leaver:1985ax}, while for the asymptotic tails the time-domain integration  \cite{Gundlach:1993tp} has been applied. The latter was used in various works as well (for instance, \cite{Konoplya:2007yy,Churilova:2019cyt,Bolokhov:2023dxq,Bronnikov:2019sbx}) with a good concordance for the dominant frequencies. As all of these methods are broadly discussed in the literature, we will not discuss them here in detail.

Using the first order WKB approach and expanding in terms of $1/L$ and $l_0$, where $L=\ell + \frac{1}{2}$ we find the position of the maximum of the effective potential:
\begin{equation}
r_{max} = 3 M-\frac{5
   l_{0}^2}{6 M} -\frac{65 l_{0}^4}{216 M^3} + \mathcal{O}(l_{0}^6),
\end{equation}
and the frequency
\begin{equation}
\begin{split}
\omega = \frac{L}{3\sqrt{3} M}-\frac{i (2n+1)}{6 \sqrt{3} M} &+ l_{0}^2 \left(\frac{L}{18 \sqrt{3} M^3}+\frac{i (2 n+1)}{54 \sqrt{3} M^3}\right) \\
&+ l_{0}^4 \left(\frac{17 L}{648 \sqrt{3} M^5}+\frac{7 i (2n+1)}{324 \sqrt{3} M^5}\right)+ \mathcal{O} \left(\frac{1}{L}, l_{0}^6\right).
\end{split}
\end{equation}

Notice, that in the eikonal limit the WKB formula is exact and the above formula satisfies the eikonal QNMs/null geodesics correspondence \cite{Cardoso:2008bp}, though in general the latter should be treated carefully, since there are a number of exceptions from it \cite{Konoplya:2017wot,Konoplya:2022gjp,Konoplya:2019hml}. 

\begin{figure}
	\begin{center}
		\begin{tabular}{cc}
			\includegraphics[width=2.3in]{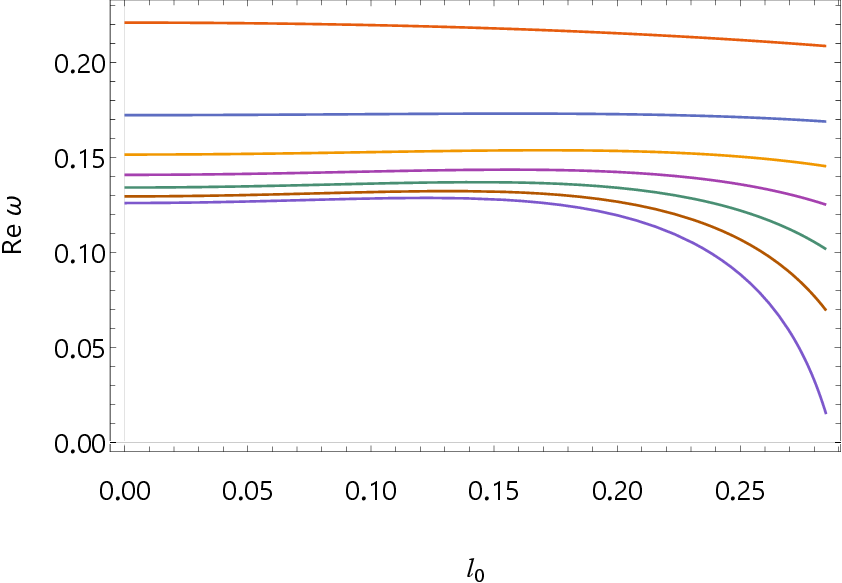} & \includegraphics[width=2.3in]{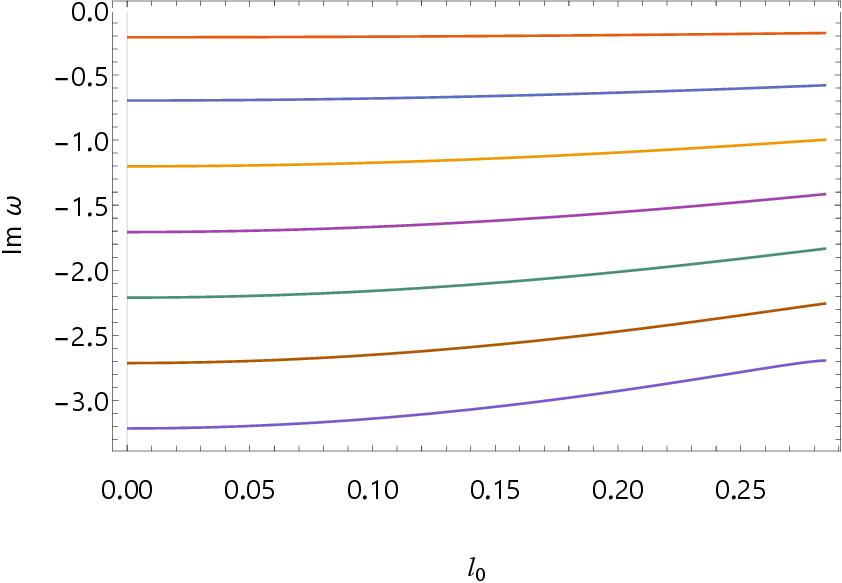}\\
		\end{tabular}
		\caption{Quasinormal frequencies for the scalar perturbations ($\ell=0$) and $n$ from $0$ to $6$ \cite{Konoplya:2023ahd}.}
		\label{fig:qnm_overtones_0n0}
	\end{center}
\end{figure}

From fig. \ref{fig:qnm_overtones_0n0} we notice that the overtones deviate from their Schwarzschild limits at an increasing with $n$ rate, which reflects the fact that the Schwarzschild metric is deformed by the $l_0$ quantum correction mainly near the event horizon \cite{Konoplya:2022pbc}. 

When the scalar field has non-zero mass $\mu$, the spectrum of the Schwarzschild and Reissner-Nordstrom black holes contains arbitrarily long lived quasinormal modes at some values of $\mu$ \cite{Ohashi:2004wr}. When $\mu$ is increased, the damping rate decreases, approaching zero as a kind of the threshold at which the mode disappear from the spectrum and the first overtone becomes the fundamental mode. This way at particular values of the mass of the field, there exist the modes, called {\it quasi-resonances},  which are similar to standing waves. In \cite{Bolokhov:2023ruj} it was shown that this phenomenon takes place also for the massive scalar field in the Bardeen background and that the outburst of overtones takes place for such modes as well.  

\section{Telling Oscillatory Tails of the Bardeen Spacetime}\label{tails}

\begin{figure*}
\includegraphics[width=3.6in]{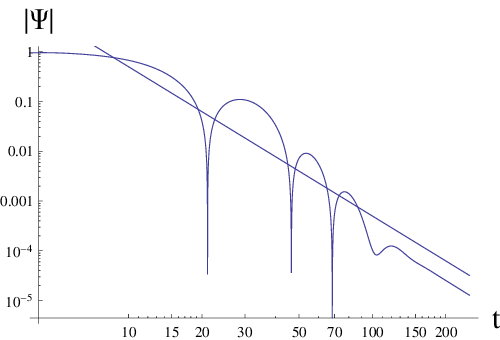}
\caption{Time-domain profile according to \cite{Konoplya:2023ahd} for scalar field perturbations $\ell=0$ in the background of the quasi-extremal Bardeen black hole $l_{0} = 0.707107$; $r_{0} =1$. Logarithmic plot with the line $\sim t^{-3}$ .}
\label{fig:TD}
\end{figure*}

At asymptotically late times the massless scalar and gravitational fields for the Schwarzschild spacetime decay according to the following law \cite{Price:1972pw}:
\begin{equation}
|\Psi| \sim t^{-(2\ell +3)}, \quad t \rightarrow \infty.
\end{equation}
In fig. \ref{fig:TD} we can see that the same law is fulfilled for the Bardeen spacetime.  

When the massive term $\mu$ is turned on, the late time behavior of the Reissner-Nordstrom black hole has two regimes \cite{Koyama:2001qw}. At {\it asymptotic} times
\begin{equation}
\frac{t}{M} > (\mu M)^{-3},
\end{equation}
the following decay law dominates
\begin{equation}
|\Psi| \sim t^{-5/6} \sin (\mu t), \quad t \rightarrow \infty.
\end{equation}
For the Bardeen spacetime, as was shown in \cite{Bolokhov:2023ruj}, the decay law is different:
\begin{equation}
|\Psi| \sim t^{-1} \sin (A(\mu) t), \quad t \rightarrow \infty,
\end{equation}
where $A(\mu)$ is some function which could be approximately found by fitting the data for various values of $\mu$.
At the {\it intermediate late} times, corresponding to relatively small value of $\mu M$, the decay law for the Bardeen spacetime is \cite{Bolokhov:2023ruj}, 
\begin{equation}
|\Psi| \sim t^{-(\frac{8}{6}+\ell)} \sin (A(\mu) t),
\end{equation}
which is also different from the Schwarzschild or Reissner-Nordstrom case \cite{Koyama:2001qw,Konoplya:2011qq}.

\section{Conclusions}\label{conclus}%%%%%%%%%%%%%%%%%%%%%%%%%%%%%%%%%%%%%%%

We have reviewed recent studies \cite{Konoplya:2023ahd,Bolokhov:2023ruj} of quasinormal modes and evolution of perturbations of a test scalar, electromagnetic and Dirac fields in the vicinity of the Bardeen spacetime treated as a quantum corrected neutral black hole \cite{Nicolini:2019irw}.  The spectrum has a number of interesting and distinctive properties such as outburst of overtones, long-lived quasinormal modes and different tail behavior at asymptotic and intermediate  times.

% Acknowledgements are created using the command \ack:
\ack%%%%%%%%%%%%%%%%%%%%%%%%%%%%%%%%%%%%%%%%%%%%%%%%%%%%%%%%%%%%%%%%%%%%%%%

% Contributors involved in `Vyzkumny zamer' can use macro \InstResCode
Authors would like to acknowledge useful discussions with A. Zhidenko and S. Bolokhov. This work was supported by the internal grant of the Silesian University in Opava SGS/30/2023.

% Here we specify the basename of the bibliography database file,
% in this case \jobname=ragsamp:
\bibliography{ragsamp}

\end{document}